\documentclass[aps,prl,twocolumn,superscriptaddress,floats,floatfix]{revtex4-1}
\usepackage{amssymb,amsmath}
\usepackage{graphics,bm}
\usepackage{epstopdf}
\usepackage{color}
\usepackage{subfigure}
\usepackage{epsfig}
\usepackage{rotating}
\usepackage{float}
\usepackage{anyfontsize}
\usepackage{ulem}
\usepackage{verbatim}
\usepackage{standalone}
\def \teddy {T_{\rm eddy}}
\def \Teddy {T_{\rm eddy}}
\def \Lint {L_{\rm I}}
\def \ellI {L_{\rm I}}

\def \ff    {\bm f} 

\def \rzero {{\bf r}_0}
\def \tzero {t_0}
\def \mP {\mathcal{P}}

\def \mQ {\mathcal{Q}}

\def \tR {t_R}

\def \rr  {{\bm r}}

\def \uu  {{\bm u}}
\def \vv  {{\bm v}}

\def \xx  {{\bm x}}

\def \grad {{\bm \nabla}}
\def \lap {\nabla^2}
\def \delt {\partial_t}

\newcommand{\bra}[1]{\langle #1\rangle}

\def \Np  {\mathcal{N}_{\rm p}}

\newcommand{\Fig}[1]{Fig.~\ref{#1}}
\newcommand{\subfig}[2]{Fig.~\ref{#1}(#2)}

\usepackage{hyperref}
\usepackage[T1]{fontenc}

\begin{document}
%
\title{The spreading of viruses by airborne aerosols: lessons from a first-passage-time 
problem for tracers in turbulent flows}
\author{Akhilesh Kumar Verma}
\email{akhilesh@iisc.ac.in}
\affiliation{Centre for Condensed Matter Theory, Department of Physics, Indian Institute of Science, Bangalore 560012, India.}
\author{Akshay Bhatnagar}
\email{akshayphy@gmail.com}
\affiliation{NORDITA, KTH Royal Institute of Technology and
Stockholm University, Roslagstullsbacken 23, 10691 Stockholm, Sweden.}
\author{Dhrubaditya Mitra}
\email{dhruba.mitra@gmail.com}
\affiliation{NORDITA, KTH Royal Institute of Technology and
Stockholm University, Roslagstullsbacken 23, 10691 Stockholm, Sweden.}
\author{Rahul Pandit}
\email{rahul@iisc.ac.in}
\altaffiliation[\\ also at~]{Jawaharlal Nehru Centre For Advanced
Scientific Research, Jakkur, Bangalore, India.}
\affiliation{Centre for Condensed Matter Theory, Department of Physics, 
Indian Institute of Science, Bangalore 560012, India.} 
\date{\today}
\begin{abstract}

We study the spreading of viruses, such as  SARS-CoV-2, by airborne aerosols,
via a new first-passage-time problem for Lagrangian tracers that are
advected by a turbulent flow: By direct numerical simulations of the
three-dimensional (3D) incompressible, Navier-Stokes equation, we
obtain the time $t_R$ at which a tracer, initially at the origin of a
sphere of radius $R$, crosses the surface of the sphere \textit{for the
first time}. We obtain the probability distribution function
$\mathcal{P}(R,t_R)$ and show that it displays two qualitatively
different behaviors: (a) for $R \ll \Lint$, $\mathcal{P}(R,t_R)$ has a
power-law tail $\sim t_R^{-\alpha}$, with the exponent $\alpha = 4$ and
$\Lint$ the integral scale of the turbulent flow; (b) for $\Lint
\lesssim R $, the tail of $\mathcal{P}(R,t_R)$ decays exponentially. We
develop models that allow us to obtain these asymptotic behaviors
analytically.  We show how to use $\mathcal{P}(R,t_R)$ to develop
social-distancing guidelines for the mitigation of the spreading of
airborne aerosols with viruses such as SARS-CoV-2. 

\end{abstract}
\maketitle

\section{Introduction}
	
By 1 June 2020 (14:31 GMT) the COVID-19 coronavirus pandemic had affected $213$
countries and territories and $2$ international conveyances; the numbers of
cases and deaths were, respectively, $6,300,444$ and
$374,527$~\cite{worldometer}.  Social distancing has played an important role
in mitigation strategies that have been used in several countries to arrest the
spread of COVID-19~\cite{socialdistancing}. To optimise social-distancing
guidelines we must ask: How far, and how fast, do small respiratory droplets or
virus-bearing aerosols spread in turbulent flows? Given the ongoing COVID-19
pandemic, it is important and extremely urgent to have at least a
semi-quantitative answer to this question. SARS-CoV-2, the virus that causes
COVID-19, spreads, principally, in two different ways: (1) First: Respiratory
droplets, ejected by the sneeze or cough of a patient, fall on nearby surfaces
or persons; in this case, approximate estimates of the distance, over which
droplets are likely to travel, are
available~\cite{NYTarticle,bourouiba2014,bourouiba2020}. (2) Second:
Transmission of this virus can occur because of airborne aerosols, such as, (a)
a cloud of fine droplets, with diameters smaller than 5 micrometers, emitted by
an infected person while speaking loudly~\cite{laserlink} or (b) the SARS-CoV-2
RNA on fine, suspended particulate matter~\cite{Setti2020}. These aerosols may
remain suspended in the air for a long time. Indeed, they have been reported in
two hospitals in Wuhan~\cite{Liu2020}; and there is growing evidence that the
SARS-CoV-2 virus could also spread via airborne
aerosols~\cite{laserlink,Setti2020,Liu2020,Prather2020,Somsen2020}, typically
indoors~\cite{Indoor2020}. Other diseases can also spread because of airborne
aerosols; examples include measles~\cite{Riley1978},
chickenpox~\cite{Leclair1980}, tuberculosis~\cite{Escombe2007}, and avian
flu~\cite{Zhao2019}.  

The typical sedimentation speed for such aerosols is comparable to their
thermal speed.  Therefore, at the simplest level, it is natural to model these
aerosol particles as neutrally-buoyant Lagrangian tracers, which are advected by
the flow, but are passive, in the sense that they do not affect the flow
velocity. We can then study the spread of viruses, such as  SARS-CoV-2, via the
airborne-aerosol route, by considering the advection of such tracers by
turbulent fluid flows. There have been extensive studies of such tracers in the
fluid-dynamics literature~\cite{Falkovich2001,Toschi2009}; and models for such
tracers have been used, \textit{inter alia}, to model the dispersion of
pheromones by lepidoptera~\cite{Celani2014}.  

\begin{figure}
\includegraphics[width=0.95\columnwidth]{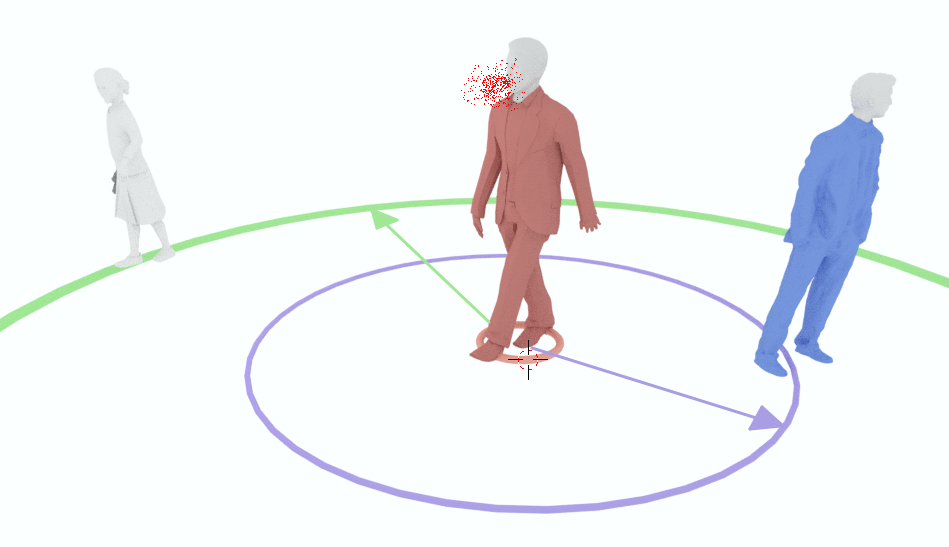}
\caption{\label{fig:SDS} A schematic diagram illustrating how aerosols with viruses
(red points) may be advected by a turbulent flow, from the person at the centre 
(the source) to other persons at different distances from the centre.} 
\end{figure}

We would like to determine the time that an aerosol particle (one of the red
particles in the schematic diagram of Fig.~\ref{fig:SDS}) takes to travel a
distance $R$ from its source (the man at the centre of Fig.~\ref{fig:SDS}). In
a turbulent flow, this time is random; furthermore, a tracer particle can go
past the distance $R$, turn back, and reach $R$ again. It is important,
therefore, to calculate the time it takes for an aerosol particle to reach the
distance $R$ \textit{for the first time} and to calculate the probability
distribution function (PDF) of the \textit{first-passage time} of a tracer in a
turbulent flow. We carry out this calculation below.

Specifically, we consider Lagrangian tracer particles that emanate from a point
source in a turbulent fluid. If $t_R$ is the time at which a tracer, initially
at the origin of a sphere of radius $R$, crosses the surface of the sphere
\textit{for the first time}, what is the probability distribution function
(PDF) $\mathcal{P}(R,t_R$)? The answer to this question is of central
importance in both fundamental nonequilibrium statistical mechanics
~\cite{bray2013persistence,chandrasekhar1943stochastic,redner2001guide,balakrishnan2008elements,
ralf2014first} and in understanding the dispersal of tracers by a turbulent
flow, a problem whose significance cannot be overemphasized, for it is of
relevance to the advection of (a) airborne viruses, as we have noted above, and
(b) pollutants in the atmosphere. First-passage-time problems have been studied
extensively~\cite{chandrasekhar1943stochastic,redner2001guide,
balakrishnan2008elements,ralf2014first} and they have found applications in a
variety of areas in physics and astronomy, chemistry~\cite{weiss1967first},
biology~\cite{ricciardi1999outline}, and finance~\cite{chicheportiche2014some}.
In the fluid-turbulence context, different groups have studied zero crossings
of velocity fluctuations~\cite{kailasnath1993zero} or various statistical
measures of two-particle dispersion, including exit-time statistics for such
dispersion in two- and three-dimensional (2D and 3D) turbulent
flows~\cite{boffetta2002statistics,vulpiani2001exit}. In contrast to these
earlier studies (e.g.,
Refs.~\cite{boffetta2002statistics,vulpiani2001exit,lalescu2018tracer}), the
first-passage-time problem we pose considers one tracer in a turbulent flow
that is statistically homogeneous and isotropic. To the best of our knowledge,
this first-passage problem has not been studied hitherto. For such a particle
we show, via extensive  direct numerical simulations (DNSs), that
$\mathcal{P}(R,t_R)$ displays a crossover between two qualitatively different
behaviors: (a) for $R \ll \Lint$, $\mathcal{P}(R,t_R) \sim t_R^{-\alpha}$, with
$\Lint$ the integral scale of the turbulent flow and the exponent $\alpha = 4$;
(b) for $\Lint \lesssim R $, $\mathcal{P}(R,t_R)$ has an exponentially decaying
tail (\Fig{fig:Qturb}). We develop models that allow us to obtain these two
asymptotic behaviors analytically. Most important, we show how to use
$\mathcal{P}(R,t_R)$ to obtain estimates of social-distancing guidelines for
the mitigation of the spreading of airborne aerosols with viruses such as
SARS-CoV-2. 

\section{Models, Methods, and Results}

\begin{figure*}
\centering	
\includegraphics[width=0.95\columnwidth]{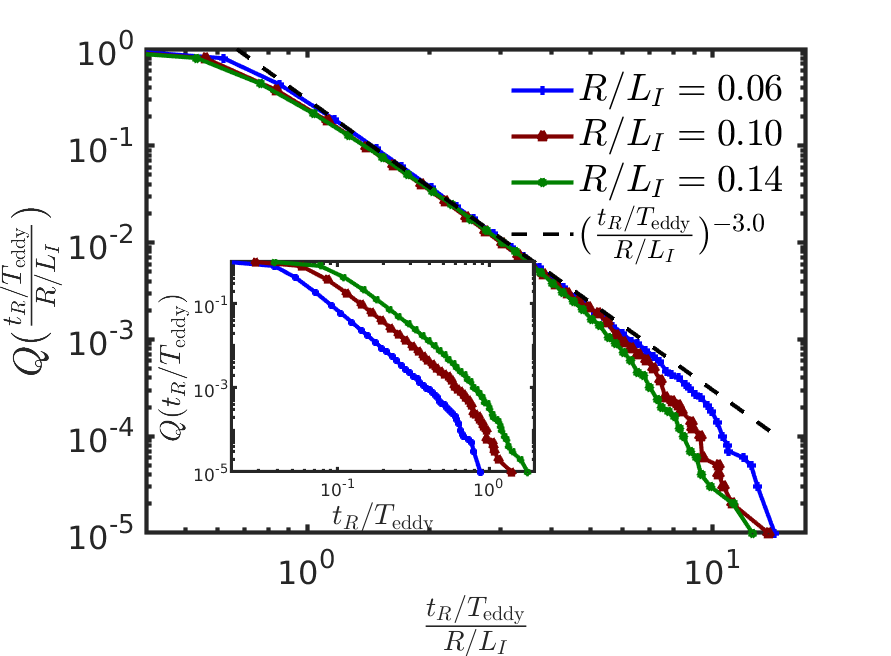}\put(-100,150){\bf (a)}\hspace{0.5cm}
\includegraphics[width=0.95\columnwidth]{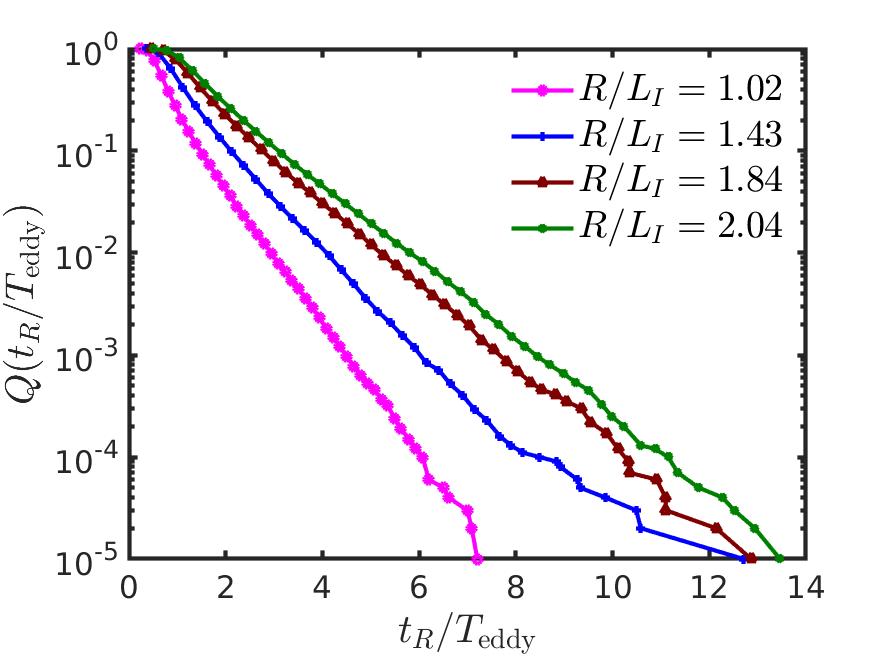}\put(-140,150){\bf (b)}
\caption{\label{fig:Qturb} Plots of the complementary cumulative probability distribution 
functions (CPDFs) $\mathcal{Q}$ versus the scaled first-passage time $\tR$ 
(see text):
(a) Log-log plots of $\mathcal{Q}(\frac{\tR/\teddy}{R/\Lint})$ for $R/\Lint = 0.06$ (blue), 
$R/\Lint=0.10$ (purple), $R/\Lint=0.14$ (green), and $(\frac{\tR/\teddy}{R/\Lint})^{-3}$ 
(black dashed line); the inset shows log-log plots of $\mathcal{Q}(\tR/\teddy)$ for
the same values of $R/\Lint$. (b) Semi-log plots of $\mathcal{Q}(\tR/\teddy)$
for $R/\Lint=1.02$ (pink), $R/\Lint=1.43$ (blue), $R/\Lint=1.84$ (purple), and 
$R/\Lint=2.04$ 
(green).}
\end{figure*}

The 3D incompressible, Navier-Stokes equation is
\begin{subequations}
\begin{equation}
\delt \uu +\left(\uu\cdot\grad\right)\uu = 
-\grad p + \nu \lap \uu + \ff \/,                
\end{equation}
and 
\begin{equation}
\grad \cdot \uu = 0\/.
\end{equation}
\label{eq:NSE}
\end{subequations}
Here, $\uu(\xx,t)$ is the Eulerian velocity at position $\xx$ at time $t$,
$p(\xx,t)$ is the pressure field, and $\nu$ is the kinematic viscosity of the
fluid; the constant density is chosen to be unity. Our direct numerical
simulation (DNS) uses the  pseudo-spectral method~\cite{canuto1988spectral},
with the $2/3$ rule for dealiasing, in a triply periodic cubical domain with
$N^3$ collocation points; we employ the second-order, exponential,
Adams-Bashforth scheme for time stepping~\cite{bhatnagar2016long}. We obtain a
nonequilibrium, statistically stationary turbulent state via a forcing term
$\ff$, which imposes a constant rate of energy
injection~\cite{lamorgese2005direct,sahoo2011systematics}, in wave-number
shells $k=1$ and $k=2$ in Fourier space; this turbulent state is statistically
homogeneous and isotropic.

To obtain the statistical properties of Lagrangian tracers, which are advected 
by this turbulent flow, we seed the flow with $\Np$ independent, identical, tracer 
particles. If the Lagrangian displacement of a tracer, which was at position 
$\rzero$ at time  $\tzero$, is $\rr(t\mid \rzero,\tzero)$, then its temporal evolution 
is given by
\begin{equation}
  \frac{d}{dt}\rr = \vv(t\mid\rzero,\tzero) = \uu(\rr,t)\/,
\label{eq:tracer}
\end{equation}
where $\vv$ is its Lagrangian velocity. In Eq.~(\ref{eq:tracer}), we need the
Eulerian flow velocity at off-grid points; we obtain this by tri-linear
interpolation; and we use the first-order Euler method for time marching (see,
e.g., Ref.~\cite{bhatnagar2016long}). We give important parameters for our DNS
runs in Table~\ref{tab:param}. These include the time step $dt$, the energy
dissipation rate $\epsilon = 2\nu \sum_k k^2E(k)$, where ${E}(k) = \sum_{k-1/2
< k' < k + 1/2} \uu({\bf k}')\cdot\uu({-\bf k}')$ is the energy spectrum, the
Taylor-microscale  $\lambda = \sqrt{\frac{2\nu E}{\epsilon}}$, where the total
energy $E = \sum_k E(k)$, the Taylor-microscale Reynolds number $Re_{\lambda}
\equiv \lambda u_{\rm rms}/\nu$,  where $u_{\rm rms} = \sqrt{2 E}$ is the
root-mean-square velocity of the flow; $\Lint = \frac{\sum_k E(k)/k}{\sum_k
E(k)}$ is the integral length scale and ${\Teddy} = \Lint/u_{\rm rms}$ is the
integral-scale eddy-turnover time; $\eta=(\nu^3/\epsilon)^{1/4}$ and
$\tau_{\eta} = (\nu/\epsilon)^{1/2}$ are, respectively, the Kolmogorov
dissipation length and time scale; and $k_{\rm {max}}$ is the maximum wave
number that we use in our DNS.

\begin{table*}
\begin{tabular}{c c c c c c c c c c c c c}
\hline
$N$ & $\nu$ & $dt$ & $Re_{\lambda}$ & $\epsilon$ & $\eta$ 
& $k_{\rm {max}}\eta$  & $\lambda$ & {$\Lint$} & {$\Teddy$} & $\tau_{\eta}$ & {$\Np$} \\
\hline
$512$ & $1.2\times10^{-3}$ & $2\times10^{-4}$ & $82$ & $0.67$ &
$7.12\times10^{-3}$ & $1.21$ & $0.08$ & $0.49$ & $0.43$ & 
$4.23\times10^{-2}$ & $100000$\\
\hline

\end{tabular}
\caption{Parameters (see text for definitions) for our DNS runs: $N^3$ is the total 
number of collocation points; $\nu$ is the kinematic viscosity; $dt$ is the time step; 
$Re_{\lambda}$ is the Taylor-microscale Reynolds number; $\epsilon$  is the 
energy-dissipation rate; $\eta$ is the Kolmogorov dissipation length scale; 
$k_{\rm {max}}$ is the maximum wave
number that we use; $\Lint$ is the integral length scale; $\Teddy$ is the
integral-scale eddy-turnover time; $\tau_{\eta} = (\nu/\epsilon)^{1/2}$ is the Kolmogorov
dissipation time scale; and $\Np$ is the number of tracer particles.}
\label{tab:param}
\end{table*}

Clearly, $\tR$ is the \textit{first} time at which $|\rr|$ becomes equal
to $R$.  Instead of computing the PDF (or histogram) of $\tR$ numerically, we
calculate the complementary cumulative probability distribution function (CPDF)
$\mathcal{Q}(\tR)$, by using the rank-order method~\cite{mit+bec+pan+fri05}, to
circumvent binning errors. In \Fig{fig:Qturb}, we present log-log and semi-log
plots of $\mathcal{Q}(\tR)$ versus $\tR/\Teddy$, for several values of $R$.
From \Fig{fig:Qturb} (a) we conclude that, for $R \ll \ellI$,
$\mathcal{Q}(\tR/\Teddy) \sim (\tR/\Teddy)^{-\alpha+1}$,  for large $\tR/\Teddy$,
with $\alpha \simeq 4$; note that, in this power-law scaling regime, the complementary 
CPDFs for different values of $R/\Lint$ collapse onto a \textit{universal scaling
form}, if we plot $\mathcal{Q}(\frac{\tR/\teddy}{R/\Lint})$.  In contrast,
\Fig{fig:Qturb} (b) shows that, for $\ellI \lesssim R$, the tail of
$\mathcal{Q}(\tR/\Teddy)$ decays exponentially. For the first-passage-time PDF,
these results imply that
\begin{equation}
  \mP(R,\tR/\Teddy) \sim
  \begin{cases}
    (\tR/\Teddy)^{-4} \quad{\rm for} \quad R \ll \Lint ; \\
    \exp(-(\tR/\Teddy)) \quad{\rm for} \quad \Lint \lesssim R . \\
    \end{cases} 
\label{eq:pdfT}
\end{equation}
We now develop models that allow us to understand these two asymptotic
behaviors analytically.

For the power-law behavior of $\mP(R,\tR)$, in the range $ R \ll \Lint$,
we construct the following, \textit{natural}, ballistic model: Tracer particles
emanate from the origin with (a) a velocity whose magnitude $v$ is a random
variable with a PDF $p(v)$; and (b) when it starts out from the origin, the
tracer's velocity vector points in a random direction. Tracers move
ballistically, for short times. Therefore, for $R \ll \Lint$, the
first-passage time $\tR = R/v$; and the first-passage PDF is
\begin{equation}
\mP(R,\tR) = \int \delta(\tR-R/v) p(v) dv.
\label{eq:FPT1}
\end{equation}
In statistically homogeneous and isotropic and incompressible-fluid turbulence, each 
component of the Eulerian velocity has a PDF that is very close to 
Gaussian~\cite{pramanareview}, so $p(v)$ has the Maxwellian~\cite{gotoh2002velocity}
form 
\begin{equation}
p(v) = C_d v^{d-1} \exp(-v^2/\sigma^2), 
\label{eq:FPT1A}
\end{equation}
where $C_d$ depends on the spatial dimension $d$, and $C_d = 4\pi$ for $d = 3$,
and $\sigma = \sqrt{\bra{v^2}} = u_{\rm rms}$.
We substitute Eq.~(\ref{eq:FPT1A}) in Eq.~(\ref{eq:FPT1}); then, by integrating over $v$, we obtain 
\begin{equation}
\mP(R,\tR) = C_d \frac{R^3}{{\tR}^{d+1}}\exp(-R^2/({\tR}^2\sigma^2)).
\label{eq:FPT1B}
\end{equation}
Therefore, in the limit of small $R$ and large $\tR$, the first-passage-time 
probability  is 
\begin{equation}
\mP(R,\tR) \sim  R^3/{\tR}^4,\hspace{0.5cm} {\rm for} \hspace{0.5cm} d=3; 
\label{eq:FPT1C}
\end{equation}
this power-law exponent is the same as the one we have obtained from our DNSs above 
(Table~\ref{tab:param} and \Fig{fig:Qturb}).

We can obtain the tail $\mP(R,\tR/\Teddy) \sim \exp(-(\tR/\Teddy))$ for $ \Lint
\lesssim R$ as follows. At times that are larger than the typical
auto-correlation time of velocities in the Lagrangian description, we follow
Taylor~\cite{tay22} and assume that the motion of a tracer particle is
diffusive. Therefore, we consider a Brownian particle in three dimensions (3D).
To calculate the first-passage-time PDF, we must first obtain the survival
probability $S(t,R|0)$, i.e., the probability that the particle has not reached
the surface of the sphere of radius $R$ up to time $t$, if it has started from
the origin of this sphere.  We start with the forward Fokker-Planck
equation~\cite{balakrishnan2008elements,Risken} for the PDF of finding the
particle at a distance $r$ from the origin at time $t$:
\begin{equation}
\frac{\partial P(r,t)}{\partial t} = K \bigg(\frac{\partial^2}{\partial r^2}+
\frac{2}{r}\frac{\partial}{\partial r}\bigg) P(r,t) \/,
\label{eq:DE1}
\end{equation}
where $K$ is the diffusion constant; this PDF satisfies the initial condition,
$P(r,0) = \delta(r)/(4\pi r^2)$ and the absorbing boundary condition
$P(R,t) = 0 $, for all $t$ at $r=R$. We obtain the following solution:
\begin{equation}
P(r,t) = \frac{1}{2R^2}\sum_{n=0}^{\infty} \frac{n}{r} \sin\bigg(\frac{n\pi r}
{R} \bigg) \exp\bigg(-K n^2 \pi^{2} t /R^2 \bigg) ,
\label{eq:DE2}
\end{equation}
whence we get
\begin{align}
S(R,t_R) & = \int_0^R P(r,t) 4 \pi r^2 dr \/ \nonumber \\
 &= 2 \sum_{n=0}^{\infty} (-1)^{n+1} \exp(-K n^2 \pi^2 t/R^2)\/,
\label{eq:DE3}
\end{align}
where, in the last step, we have used Eq.~(\ref{eq:DE2}). 
The first-passage-time probability is
\begin{align}
\mP(R,t_R)&=-\frac{\partial}{\partial t_R} S(R,t_R) \nonumber \\
 & = \frac{2K\pi^2}{R^2} \sum_{n=0}^{\infty} (-1)^{n+1} n^2
\exp(-K n^2 \pi^2 t_R/R^2).
\label{eq:FPT}
\end{align}
At large times, the first term ($n=1$) is the dominant one; therefore,
\begin{equation}
\mP(R,t_R) \sim (1/R^2) \exp(-K \pi^2 t_R/R^2), 
\label{eq:FPTA}
\end{equation}
the exponential form that we have obtained from our DNS
(\Fig{fig:Qturb} (b)); the $1/R^2$ pre-factor cannot be extracted
reliably from our DNS data, because this requires much longer runs 
than are possible with our computational resources.   

\begin{figure*}
\includegraphics[width=0.95\columnwidth]{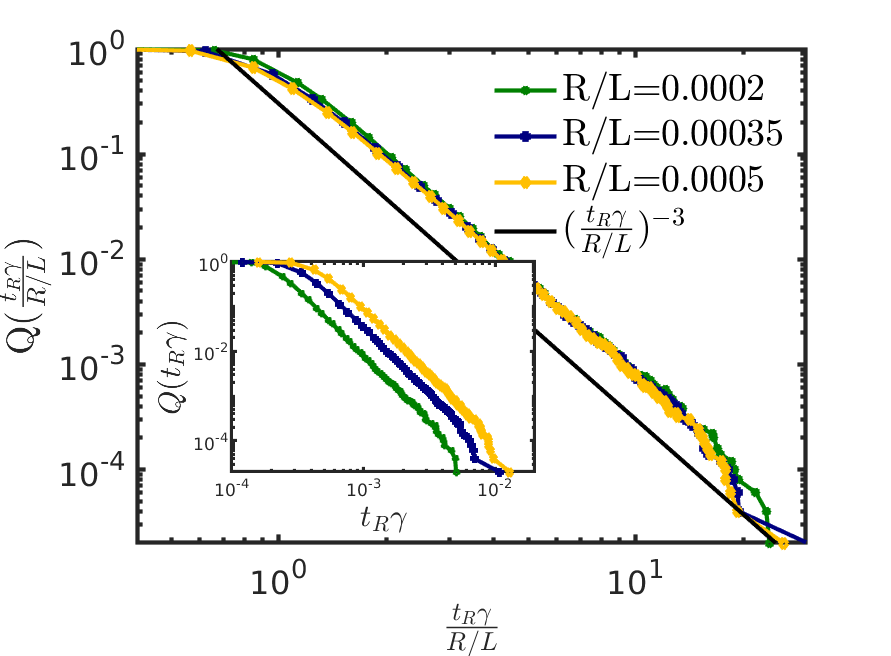}\put(-105,150){\bf (a)}\hspace{0.5cm} 
\includegraphics[width=0.95\columnwidth]{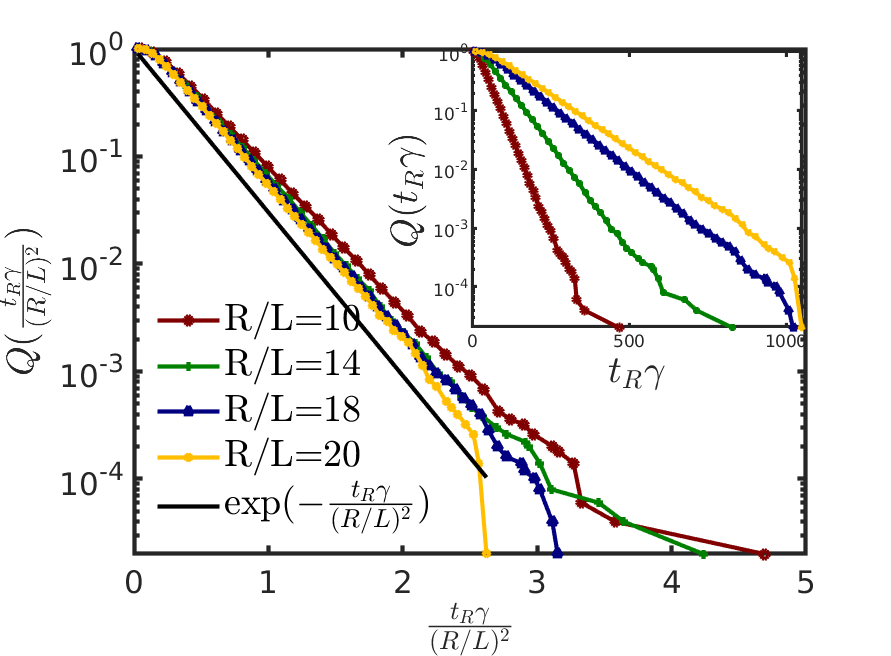}\put(-140,150){\bf (b)} 
\caption{\label{fig:QPOU}(a) Log-log plots of the complementary CPDFs 
$\mathcal{Q}(\frac{\gamma\tR}{R/L})$ of the scaled 
first-passage time $\frac{\gamma\tR}{R/L}$ , for $R \ll L$ and $\gamma = 0.01$; the complementary CPDFs, for $R/L = 0.0002$ (green), $R/L = 0.00035$ (blue), 
and $R/L = 0.0005$ (orange), collapse onto one curve; (b) semi-log plots of the
complementary CPDFs of the scaled first-passage time $\tR/R^2$, for 
$L \lesssim R$ and $\gamma = 30$.  The complementary CPDF of $\tR/R^2$, for 
$R/L = 10$ (purple), $R/L = 14$ (green), $R/L =18$ (blue), and $R/L = 20$ 
(orange), collapse onto one curve. Plots of the complementary CPDFs 
$\mathcal{Q}(\gamma\tR)$ versus $\gamma\tR$ are shown in the insets.}

\end{figure*}

We now show that both the small- and large-$R/\Lint$ behaviors of
$\mP(R,t_R)(R,t_R)$ in Eq.~(\ref{eq:pdfT}) can be obtained from one stochastic
model for the motion of a particle. The simplest such model uses a particle
that obeys the following Ornstein-Uhlenbeck (OU) model:
\begin{subequations}
  \begin{align}
    \frac{d {x}_i}{dt} &= {v_i} \/, \\ 
    \frac{d{v}_i}{dt} &= -\gamma v_i + \frac{\sqrt{\Gamma}}{m}\zeta_i \/.
  \end{align}
\label{Eq:OUE}
\end{subequations}
Here, $\gamma$ and $\Gamma$ are positive constants; $x_i$ and $v_i$ are the
Cartesian components of the position and velocity of the particle; in three
dimensions, $ i = 1, 2$, and $3$; $\zeta_i (t)$ is a zero-mean Gaussian white
noise with $\bra{\zeta_i} = 0 $ and $\bra{\zeta_i(t) \zeta_j(t^{\prime}) }
=\delta_{ij}\delta(t-t^{\prime})$; this noise is such that the
fluctuations-dissipation theorem (FDT) holds.  Note that there is no FDT for
turbulence. However, for the one-particle statistics that we consider, the
simple OU model is adequate. We use $N_p = 50,000$ particles; for each
particle, the initial-position components $x_i(t=0)$ are distributed randomly
and uniformly on the interval $[0,2\pi]$; and the velocity components
$v_i(t=0)$ are chosen from a Gaussian distribution. For each particle, we
obtain, numerically, the time $\tR$ at which it reaches a distance $R$ from the
origin \textit{for the first time}. We then obtain the first-passage-time
complementary CPDF $\mQ(\tR)$, which we plot in \Fig{fig:QPOU}, for $R \ll L$
and $L \lesssim R$, where $L = \sqrt{\frac{\Gamma}{\gamma^3}}$, the natural
length scale for Eq.~(\ref{Eq:OUE}),  plays the role of $\Lint$ in our DNSs
above (Table~\ref{tab:param} and \Fig{fig:Qturb}). We find  
\begin{eqnarray}
\mP(R,\tR) &\sim& \Bigl[\frac{\gamma\tR}{(R/L)}\Bigr]^{-4},\, {\rm for} \, R \ll L; \nonumber \\ 
\mP(R,\tR) &\sim& \exp\Bigl(-\frac{\gamma\tR}{(R/L)^2}\Bigr),\, {\rm for} \,  L \lesssim R;
\label{Eq:pdfQU}
\end{eqnarray}
these are the OU-model analogs of our DNS results Eq.~(\ref{eq:pdfT}).  We have
carried out two OU-model simulations: (a) we have designed the first, with
$\gamma = 0.01$, to explore the form of $\mP(R,t_R)$ in the ballistic regime $R
\ll L$; (b) the second, with $\gamma = 30$, allows us to uncover the form of
$\mP(R,t_R)$ in the diffusive regime $ L \lesssim R$. (From a numerical
perspective, it is expensive to obtain the precise form of $\mP(R,t_R)$ in both
ballistic and diffusive regimes, with one value of $\gamma$.) We now explore in
detail the forms of $\mP(R,t_R)$ in these two regimes.  In
\subfig{fig:QPOU}{a}, we present log-log plots of the complementary CPDFs of
the scaled first-passage time $\tR/R$, for $R \ll L$ and $\gamma = 0.01$. The
complementary CPDFs of $\tR/R$, for $R/L = 0.0002$, $R/L = 0.00035$, and $R/L =
0.0005$, collapse onto one curve; i.e., in this regime, $\tR$ scales as $R$,
which is a clear manifestation of ballistic motion. In \subfig{fig:QPOU}{b}, we
present semi-log plots of the complementary CPDFs of the scaled first-passage
time $\tR/R^2$, for $L \lesssim R$ and $\gamma = 30$.  The complementary CPDFs
of $\tR/R^2$, for $R/L = 10$, $R/L = 14$, $R/L =18$, and $R/L = 20$, collapse
onto one curve; from this we conclude that, in this regime, $\tR$ scales as
$R^2$, which is a clear signature of diffusive motion.

\section{Conclusions and Discussion}

We have defined and studied a new first-passage-time problem for Lagrangian
tracers that are advected by a 3D turbulent flow that is statistically steady,
homogeneous and isotropic. Our work shows that the first-passage-time PDF
$\mP(R,\tR)$ has tails that cross over from a power-law form to an
exponentially decaying form as we move from the regime $R \ll \Lint$ to $\Lint
\lesssim R$ (Eq.~(\ref{eq:pdfT})). We develop ballistic-transport and diffusive
models, for which we can obtain these limiting asymptotic behaviors of
$\mP(R,t_R)$ analytically.  We also demonstrate that an OU model, with Gaussian
white noise, which mimics the effects of turbulence, suffices to obtain the
crossover between these limiting forms. Of course, such a simple stochastic
model cannot be used for more complicated multifractal properties of turbulent
flows ~\cite{vulpiani2001exit,pramanareview,arneodo}.

Earlier studies have concentrated on two-particle relative dispersion by using
doubling-time statistics, in 2D fluid turbulence; in particular, they have
shown that the PDF of this doubling time has an exponential
tail~\cite{boffetta2002statistics}.  Studies of velocity zero
crossings~\cite{kailasnath1993zero}, in a turbulent boundary layer, have shown
that PDFs of the zero-crossing times have exponential tails. 

The single-particle first-passage-time statistics that we study have not been
explored so far. Furthermore, $\mP(R,\tR)$ can be used to to develop
social-distancing guidelines for the mitigation of the spreading of airborne
aerosols with viruses such as SARS-CoV-2 as we show below. 

\begin{figure*}
\includegraphics[width=0.95\columnwidth]{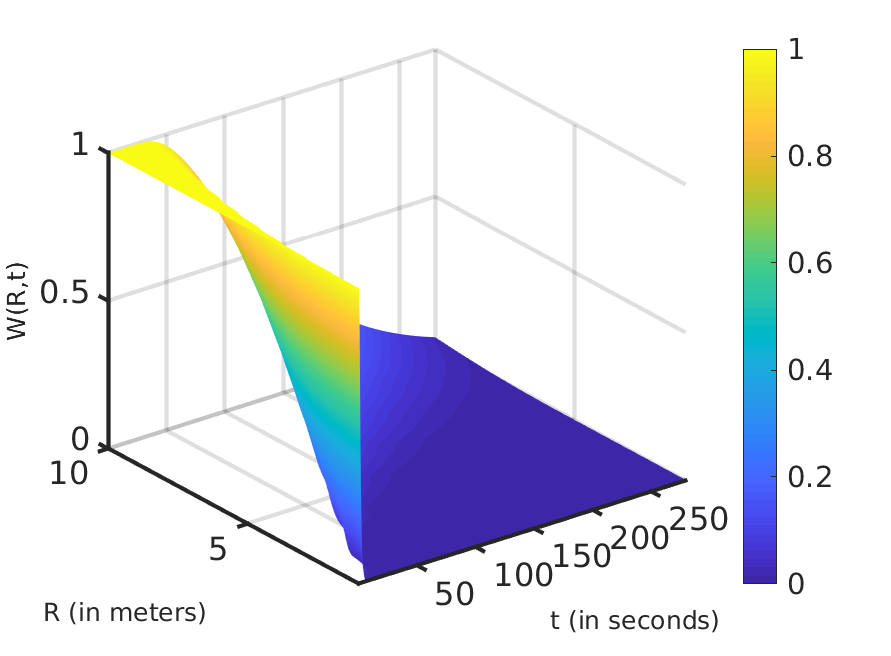}\put(-105,150){\bf (a)}\hspace{0.5cm} 
\includegraphics[width=0.95\columnwidth]{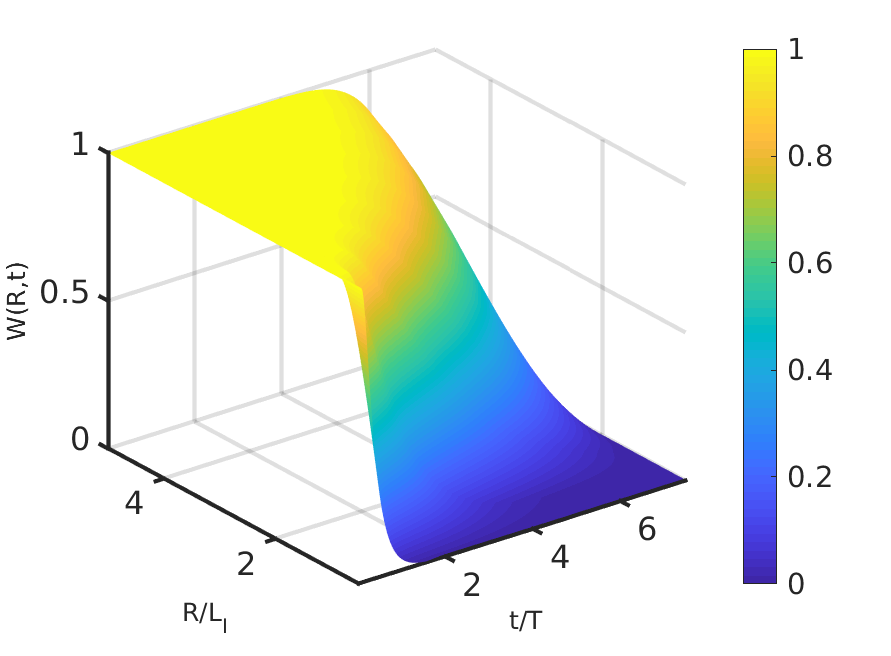}\put(-140,150){\bf (b)} 
\caption{\label{fig:SWP} Surface plots of $W(R,t)$, the
probability of virus-laden aerosol particles \textit{not reaching} a
person (at a distance $R$ from an infected person), \textit{up until
time} $t$ (the first-passage time) in (a) versus $R$ and $t$, and (b)
versus $R/L_I$ and $t/T$, in the diffusive region, for the representative
values $K=0.1 {\rm m}^2/{\rm s}$, $u_{\rm rms} = 0.05$ m/s, and $\Lint = 2 $m.}
\end{figure*}

Given a pseudospectral DNS, of the type we have carried out, we can obtain the
integral scale $\Lint$ and $u_{\rm rms}$ from the energy spectrum $E(k)$, as we
have noted above. A recent study of COVID-19 in $320$ municipalities in China
suggests that a very large fraction of COVID-19 infections occur because of
indoor transmission of the SARS-CoV-2 virus~\cite{Indoor2020}.  Therefore, it
is important to study such transmission in rooms and offices; a comprehensive
DNS study of the Navier-Stokes equation, with the correct boundary conditions
enforced at every wall and surface in the room and accurate forcing functions
(dictated, e.g., by fans and vents), is a considerable challenge. Furthermore,
it is not possible to carry out such a DNS for every room with a different
arrangement of the furniture in it.  Hence, it is important to come up with
semi-quantitative criteria that help us to understand, and mitigate, the indoor
transmission of such virues.  Turbulence models have been used to study the
flow of air in rooms and offices~\cite{li2005multi,zhang2007evaluation}; from
these models and related experiments we obtain the estimate $u_{\rm rms} \simeq
0.05$ m/s in a typical office.  We must also estimate $\Lint$, for it is an
important crossover length scale in our analysis of $\mP(R,\tR)$. In our DNS
$\Lint$ is $\simeq 0.1 L$, where $L$ is the linear size of our simulation
domain. In a typical office or a train, with fixed forcing, via fans or vents,
we use $\Lint$ to be approximately a few meters; of course, $\Lint$ must depend
on the degree of crowding on a train or the number of cubicles in a large
office room.  Now consider one infected person who is at a distance $R$ from
another person.  The probability of virus-laden aerosol particles \textit{not
reaching} the second person, \textit{up until time} $t$ is related to
$\mP(R,\tR)$ as follows:
\begin{equation}
W(R,t) = 1-\frac{\mP(R,\tR=t)}{\Sigma_{\tR} \mP(R,\tR)},
\end{equation}
which we calculate, by using  Eq.~(\ref{eq:FPT}), and depict in
Fig.~\ref{fig:SWP}(a) and Fig.~\ref{fig:SWP}(b), in the diffusive regime;
Fig.~\ref{fig:SWP}(a) is a surface plot of $W(R,t)$ versus $R$ and $t$, for the
representative values $\Lint = 2$ m and $u_{\rm rms} = 0.05$ m/s;
Fig.~\ref{fig:SWP}(b) gives a surface plot of $W$ versus the dimensionless
parameters $R/\Lint$ and $t/T$, where $T = \Lint/u_{\rm rms}$. (We give similar
plots for the ballistic regime in the Supplemental Material.) In
Table~\ref{tab:prob_data} we give the values of $W(R,t)$ for different values
of $R$ and $t$. These figures and Table~\ref{tab:prob_data} lead to three clear
observations: 
\begin{enumerate}
\item If the separation $R \ll \Lint$, i.e., we have to consider the ballistic regime
(see the Supplementary Material), then
$W(R,t)$ goes very rapidly to $0$ (i.e., the aerosol particle reaches the 
second person), even if $t$ is very small. 
\item The smaller the separation $R$,
between two persons, the shorter the time $t$ in which $W(R,t)$ becomes
very small, i.e., the aerosol particles reach from one person to the other. 
\item Our calculation leads to quantitative predictions; e.g.,
if the separation $\Lint \lesssim R$, i.e., we have to consider the diffusive regime, 
then $W(R,t)$ goes to $0$ in tens of seconds, if $R = 2$ m, and in hundreds 
of seconds, $R = 10$ m, for the representative parameters that we use to
obtain  Table~\ref{tab:prob_data}.
A recent study ~\cite{vancovid} has suggested that the SARS-CoV-2 virus remains 
viable in aerosols for nearly $3$ hours. Therefore, if the concentration of 
virus-laden aerosols is high in a poorly ventilated room, then we must employ 
more stringent social-distancing norms than are in place now, even if people
spend only tens of minutes together in such a room.
\end{enumerate}

\begin{table*}
\begin{tabular}{c c c c c c c c c c }
\hline
$ $ & $t = 10$  s & $t = 30$ s & $t= 60$ s & $t = 100$ s & 
$t = 120$ s & $t = 150$ s & $t = 180$ s & $t = 240$ s  \\
\hline
$R = 2$ m & $0.169$ & $0.001$ & $0.000$ & $0.000$ & $0.000$ & $0.000$ & $0.000$ & $0.000$  \\
\hline
$R = 3$ m & $0.642$ & $0.074$ & $0.003$ & $0.000$ & $0.000$ & $0.000$ & $0.000$ & $0.000$  \\
\hline
$R=4$ m & $0.917$ & $0.313$ & $0.049$ & $0.004$ & $0.001$ & $0.000$ & $0.000$ & $0.000$ \\
\hline
$R=5$ m & $0.989$ & $0.594$ & $0.187$ & $0.039$ & $0.018$ & $0.005$ & $0.001$ & 
$0.000$ \\
\hline
$R = 6$ m & $0.999$ & $0.805$ & $0.383$ & $0.129$ & $0.074$ & $0.032$ & $0.014$
& $0.002$ \\
\hline
$R = 7$ m & $1.000$ & $0.923$ & $0.579$ & $0.263$ & $0.174$ & $0.093$ & $0.049$
& $0.011$ \\
\hline
$R = 8$ m & $1.000$ & $0.097$ & $0.739$ & $0.412$ & $0.299$ & $0.182$ & $0.107$& $0.030$ \\
\hline
$R = 9$ m & $1.000$ & $0.099$ & $0.850$ & $0.553$ & $0.428$ & $0.283$ & $0.180$
& $0.059$ \\
\hline
$R = 10$ m & $1.000$ & $0.998$ & $0.920$ & $0.673$ & $0.547$ & $0.386$ & $0.260$& $0.093$ \\
\hline
\end{tabular}
\caption{Table of values of $W(R,t)$, the
probability of virus-laden aerosol particles \textit{not reaching} a
person (at a distance $R$ from an infected person), \textit{up until
time} $t$ (the first-passage time) in the diffusive region, for the representative 
values $K=0.1 {\rm m}^2/{\rm s}$, $u_{\rm rms} = 0.05$ m/s, and $\Lint = 2 $m.}
\label{tab:prob_data}
\end{table*}
The methods that we have developed can be applied, \textit{mutatis mutandis}, 
(a) in sophisticated models for virus particles or droplets, e.g., 
those that use inertial particles~\cite{Cencini2006,Salazar2009} or multi-phase 
flows~\cite{Balachandar2010,Pal2016} and (b) in turbulent flows that are
not statistically homogeneous and isotropic. We will examine these in future work.
At the moment, it is important to use our minimal model to obtain semi-quantitative
for social-distancing guidelines, as we have done above.

\begin{acknowledgments} 

D.M. and A.B. thank John Wettlaufer for useful discussions. A.K.V. and R.P.
thank Jaya Kumar Alageshan for discussions and for help with
Fig.~\ref{fig:SDS}, and CSIR (IN) and DST (IN) for financial support.
A.B. and D.M. are supported by the grant Bottlenecks for Particle
Growth in Turbulent Aerosols from the Knut and Alice Wallenberg
Foundation (Dnr.  KAW 2014.0048); the computations were performed on
resources provided by the Swedish National Infrastructure for Computing
(SNIC) at PDC and SERC (at IISc).

\end{acknowledgments}


\begin{center}
\pagebreak
\widetext
\textbf{\large Supplemental Materials:The spreading of viruses by airborne aerosols: lessons
from a first-passage-time problem for tracers in turbulent flows}
\author{Akhilesh Kumar Verma}
\email{akhilesh@iisc.ac.in}
\affiliation{Centre for Condensed Matter Theory, Department of Physics, Indian Institute of Science, Bangalore 560012, India.}
\author{Akshay Bhatnagar}
\email{akshayphy@gmail.com}
\affiliation{NORDITA, KTH Royal Institute of Technology and
Stockholm University, Roslagstullsbacken 23, 10691 Stockholm, Sweden.}
\author{Dhrubaditya Mitra}
\email{dhruba.mitra@gmail.com}
\affiliation{NORDITA, KTH Royal Institute of Technology and
Stockholm University, Roslagstullsbacken 23, 10691 Stockholm, Sweden.}
\author{Rahul Pandit}
\email{rahul@iisc.ac.in}
\altaffiliation[\\ also at~]{Jawaharlal Nehru Centre For Advanced
Scientific Research, Jakkur, Bangalore, India.}
\affiliation{Centre for Condensed Matter Theory, Department of Physics,
Indian Institute of Science, Bangalore 560012, India.}
\end{center}
\setcounter{equation}{0}
\setcounter{figure}{0}
\setcounter{table}{0}
\setcounter{page}{1}
\makeatletter
\renewcommand{\theequation}{S\arabic{equation}}
\renewcommand{\thefigure}{S\arabic{figure}}
\renewcommand{\bibnumfmt}[1]{[S#1]}
\renewcommand{\citenumfont}[1]{S#1}

\begin{figure*}
\resizebox{\linewidth}{!}{
\includegraphics[width=0.95\columnwidth]{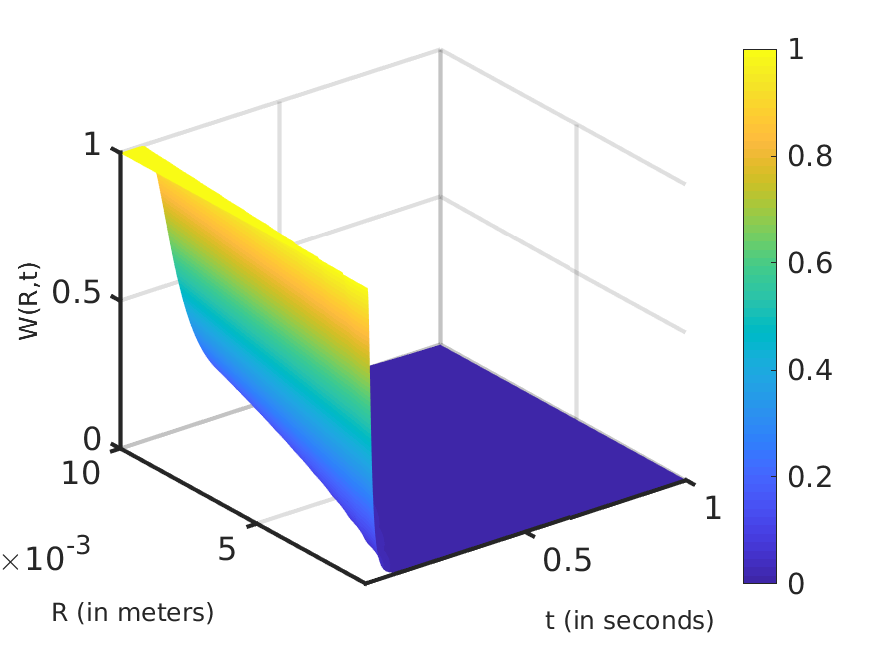}\put(-105,150){\bf (a)}\hspace{0.5cm}
\includegraphics[width=0.95\columnwidth]{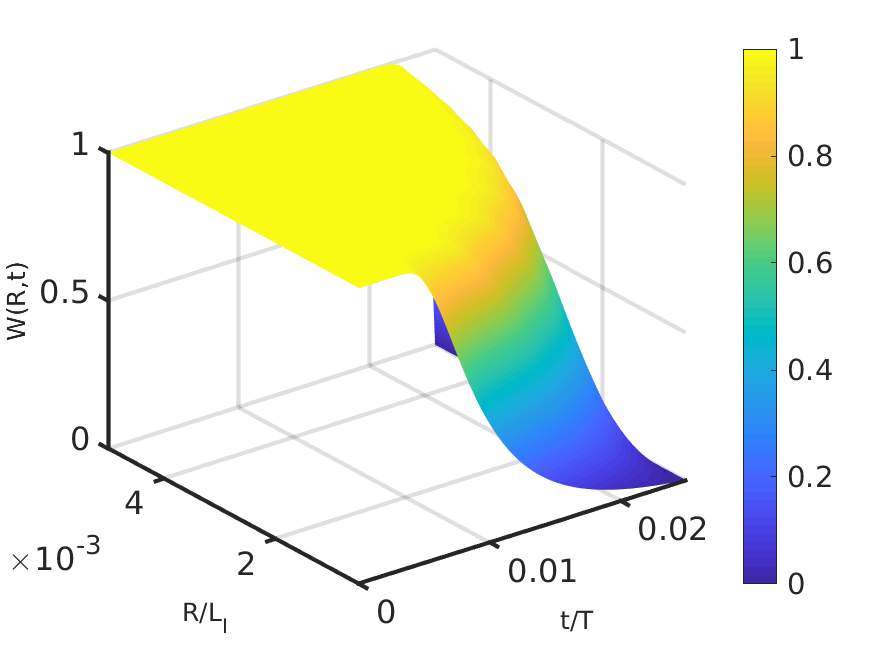}\put(-140,150){\bf (b)}}
\caption{\label{fig:NISP}  Surface plots of $W(R,t)$ versus (a) $R$ and $\tR$ and (b)
(b) the scaled radius $R/\Lint$ and scaled time $t/T$ (see the main
paper), in the ballistic region, for the representative values
$u_{\rm rms} = 0.05$ m/s, and $\Lint = 2 $m.}
\end{figure*}

\begin{figure*}
\resizebox{\linewidth}{!}{
\includegraphics[width=0.95\columnwidth]{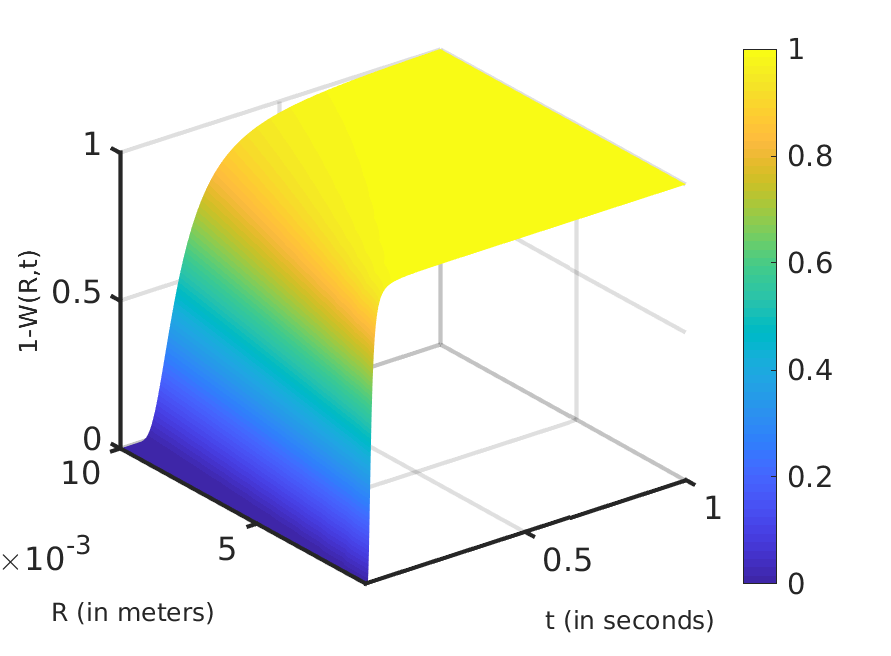}\put(-105,150){\bf (a)}\hspace{0.5cm}
\includegraphics[width=0.95\columnwidth]{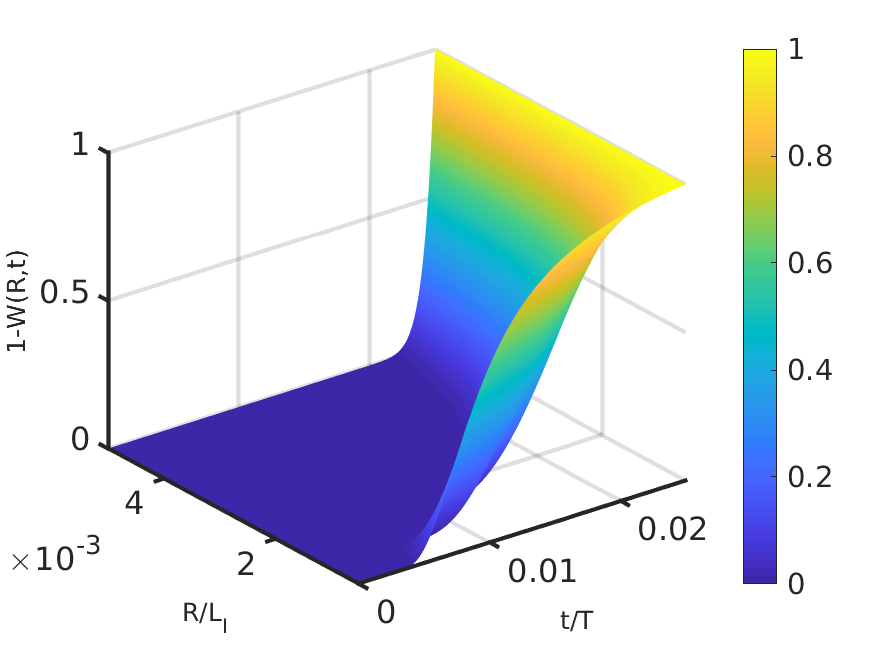}\put(-140,150){\bf (b)}}
\caption{\label{fig:ISP} Surface plots of $(1-W(R,t))$ versus (a) $R$ and $t$ and (b)
(b) the scaled radius $R/\Lint$ and scaled time $t/T$ (see the main
paper), in the ballistic region, for the representative values
$u_{\rm rms} = 0.05$ m/s, and $\Lint = 2 $m.}
\end{figure*}

\begin{figure*}
\resizebox{\linewidth}{!}{
\includegraphics[width=0.95\columnwidth]{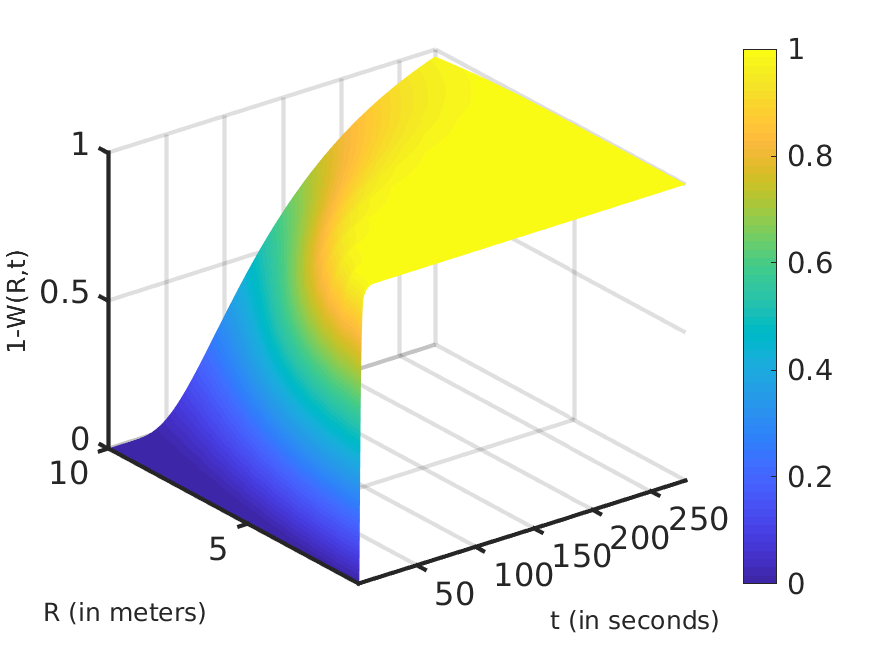}\put(-105,150){\bf (a)}\hspace{0.5cm}
\includegraphics[width=0.95\columnwidth]{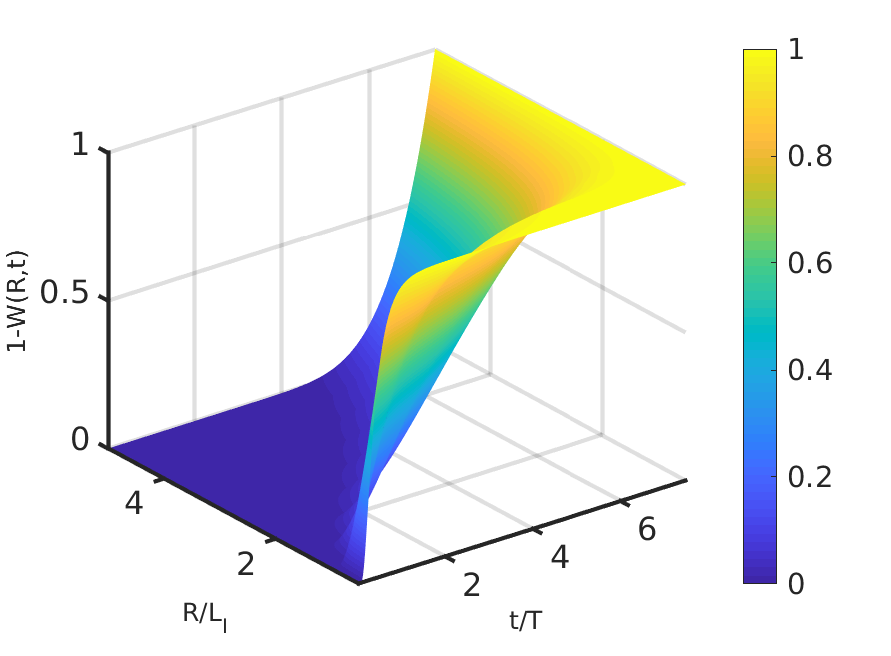}\put(-140,150){\bf (b)}}
\caption{\label{fig:DISP}  Surface plots of $(1-W(R,t))$ versus (a) $R$ and $t$ and (b)
(b) the scaled radius $R/\Lint$ and scaled time $t/T$ (see the main
paper), in the diffusive region, for the representative values
$u_{\rm rms} = 0.05$ m/s, $K=0.1 {\rm m}^2/{\rm s}$, and $L_I = 2 $ m.}
\end{figure*}

In this Supplemental Material, we provide the following:

(a) Consider one infected person, who is at a distance $R$ from another person. We give
surface plots of the probability $W(R,t)$ of a virus-laden aerosol particle
\textit{not reaching} the second person, \textit{up until time} $t$,
in the ballistic regime (Fig.~\ref{fig:NISP}). We also give surface plots of
$(1-W(R,t))$, the probability that a virus-laden aerosol particle
\textit{reaches} the second person, \textit{at time} $t$, for the first time
for ballistic (Fig.~\ref{fig:ISP}) and diffusive (Fig.~\ref{fig:DISP}) cases.

\begin{figure}
\resizebox{\linewidth}{!}{
\includegraphics[scale=0.7]{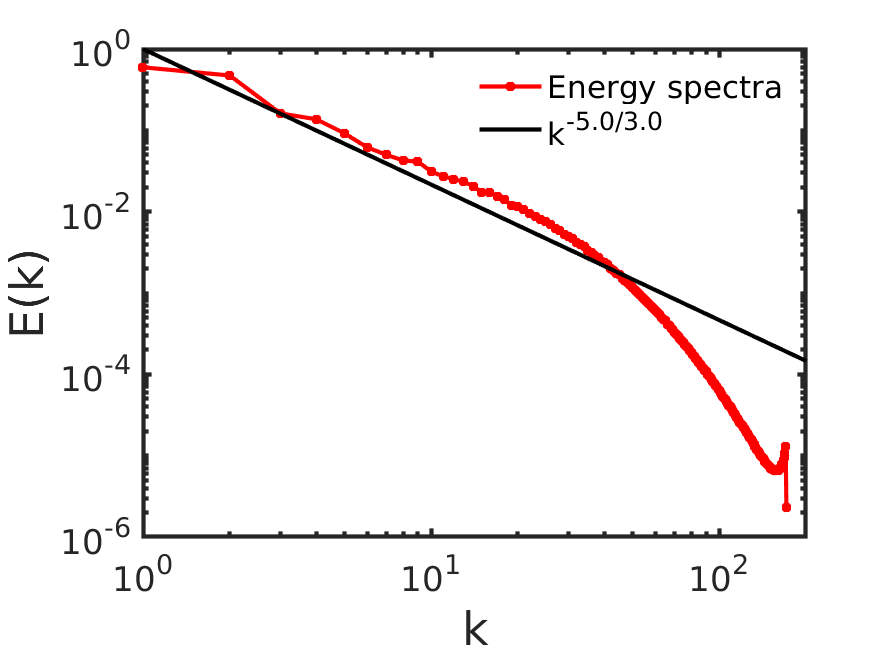}}
\caption{Log-log plot of the energy spectrum ${E}(k)$  (red curve) versus the
wave number $k$ from our direct numerical simulation of the
three-dimensional Navier-Stokes equation (see the main paper); for comparison,
we show the Kolmogorov scaling form $E(k) \sim k^{-5/3}$ (black line).}
\label{fig:ESP}
\end{figure}

%
(b) The energy spectrum ${E}(k) = \sum_{k-1/2 < k' < k + 1/2} \uu({\bf
k}')\cdot\uu({-\bf k}')$ from our direct numerical simulation of the
three-dimensional Navier-Stokes equation (see the main paper). A log-log
plot of this spectrum (red curve) is given in Fig.~\ref{fig:ESP}; for comparison,
we show the Kolmogorov scaling form $E(k) \sim k^{-5/3}$ (black line).

\end{document}